\documentclass[a4paper]{article}
\usepackage[latin1]{inputenc}
\usepackage[dvips]{graphics}
\usepackage[small]{caption}

\title{Molecular Phylogenetic Analyses and Real Life Data}
\author{Kerstin Hoef-Emden}

\begin{document}

\maketitle

\begin{center}

Universität zu Köln, Botanisches Institut, Lehrstuhl I, Gyrhofstr. 15,

50931 Köln, Germany

e-mail: kerstin.hoef-emden@uni-koeln.de

\end{center}

\bigskip

\section{What is Molecular Phylogeny?}

Most probably, all life existing today on earth shares a common ancestry
billions of years back in the past. A set of indispensable genes necessary
for maintenance of basic cell functions were passed on from the unknown
common ancestor to its extant descendants by asexual and/or sexual
reproduction. During the course of evolution, the genes, the numbers of
genes, their functions and the sizes of the genomes (i.e.\ the total DNA
content of a cell) became modified. If genes originate from a common
ancestor gene and fulfill the same function in a cell, they are said to be
homologous. The degree of divergence between homologous genes is considered
a measure for their relatedness (and also for the relatedness of the
organisms).

In molecular phylogeny, the relationships among, usually extant, organisms
are examined by comparing homologous DNA or protein sequences (i.e.\ the gene
products). The relationships are displayed as trees with branch (or edge)
lengths reflecting the degrees of genetic divergence. Each branch tip
represents an extant sequence; the internal nodes or vertices represent
unknown ancestors to the terminal nodes. The branching pattern and branch
lengths describe the evolutionary pathways leading to the sequences at the
terminal nodes. Clusters of terminal branches connected to a common ancestor
are termed clades.

The construction of phylogenetic trees has been shown to be a NP-hard
problem; the number of possible trees increases exponentially with the
number of DNA or protein sequences included in the phylogenetic analyses
\cite{Steel1992}. Due to the large amount of data and the complexity of the
task, phylogenetic trees cannot be inferred without help of computers.

Numerous studies addressing the problems of molecular phylogenetic analyses
methods in theory or practice have been published. First publications about
phylogenetic methods date back into the 60s. The methods and evolutionary
models were refined in the course of time, but problems still remain. The
cited references in this review represent only few examples from a vast
amount of literature. Also only some of the mostly used methods in molecular
phylogeny are presented.

For digging into the mathematics behind the phylogenetic analyses methods
introduced below, one may start with Joe Felsenstein's book
\cite{Felsenstein2003}.

\section{Phylogenetic Analyses Methods}

DNA sequences are based on a four-letter-code representing the four
nucleotides (A for adenin, C for cytosin, G for guanin, T for thymin),
whereas protein sequences are based on a twenty-letter-code representing the
twenty different amino acids. Prior to the phylogenetic analyses, an
alignment of the sequences has to be assembled (the single sequence is also
termed a ``taxon", because it represents a species, genus, individual or
strain). If sequences of homologous genes e.g.\ show differences in lengths
due to insertions or deletions, gaps have to be inserted to place
functionally corresponding positions in the same vertical column of the
alignment (Fig.\ 1).
\begin{figure}[h] 
 \begin{center}
  \includegraphics{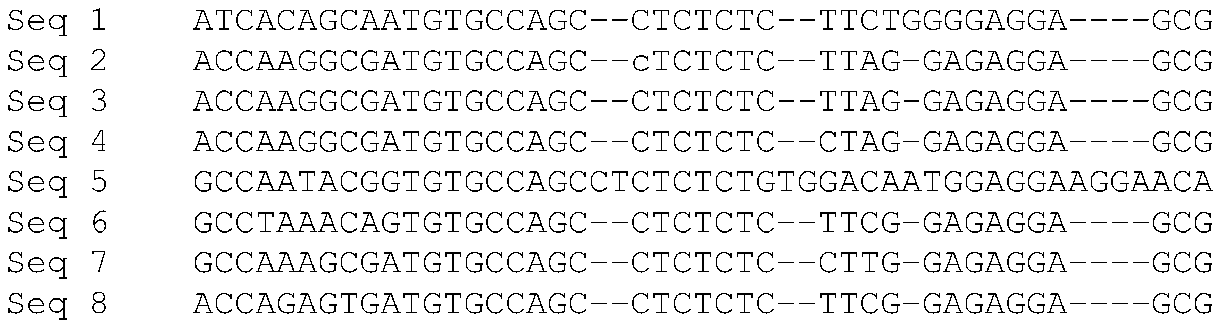}
  \caption{Excerpt from an alignment of nuclear ITS2 sequences. The ITS2 or
internal transcribed spacer 2 expands between two RNA coding genes of the
ribosomal operon. The ribosomal operon is transcribed in one piece. The two
internal transcribed spacers between the RNA coding regions fold up in a
specific way and are excised. Since the two ITS regions solely function as
spacers, they are under low selective pressure and, thus, display high
mutation rates. The example alignment shows ITS2 regions of closely related
organisms belonging to one genus. The sequences are oriented in horizontal
direction, whereas functionally corresponding positions are arranged in
columns. Several gaps had to be inserted due to insertions of nucleotides in
the sequences 1 and 5.}
 \end{center}
\end{figure}
Non-alignable regions such as insertions of several nucleotides need to be
excluded from the phylogenetic analyses. Improperly aligned sequences or
inclusions of non-alignable regions in the phylogenetic analyses may result
in artefactual phylogenetic trees.

In most standard methods for inferring phylogenetic trees, an optimality
criterion and a tree search algorithm have to be chosen. The optimality
criterion is used to determine the best among the considered trees by
defining a type of ``scoring" system. Optimality criteria are e.g.\ maximum
parsimony, distance matrix or maximum likelihood \cite{Felsenstein2003}.

In unweighted maximum parsimony, each mutation from one nucleotide or amino
acid to another, e.g.\ from a C to a G, costs one ``penalty" point. All point
mutations are considered equally likely. The mutations along a given tree
are summed up and the best tree or maximum parsimony tree is the one with
the lowest sum of penalty points. Unweighted maximum parsimony uses integer
values and often several to many equally parsimonious trees are found.

In distance analyses, the sequences are pair-wise compared. Their genetic
divergences are transformed into distance values and listed in a triangular
distance matrix. Whereas maximum parsimony treats all mutations as equally
likely, the computation of distance matrices allows for different mutation
rates and other variations of parameters (i.e.\ evolutionary models, see
chapter below). To infer trees from a distance matrix, usually the
neighbor-joining algorithm is used (see below).

Maximum likelihood is a probablistic and the computationally most costly
method (Fig.\ 2).
\begin{figure}[htbp]
 \begin{center}
  \includegraphics{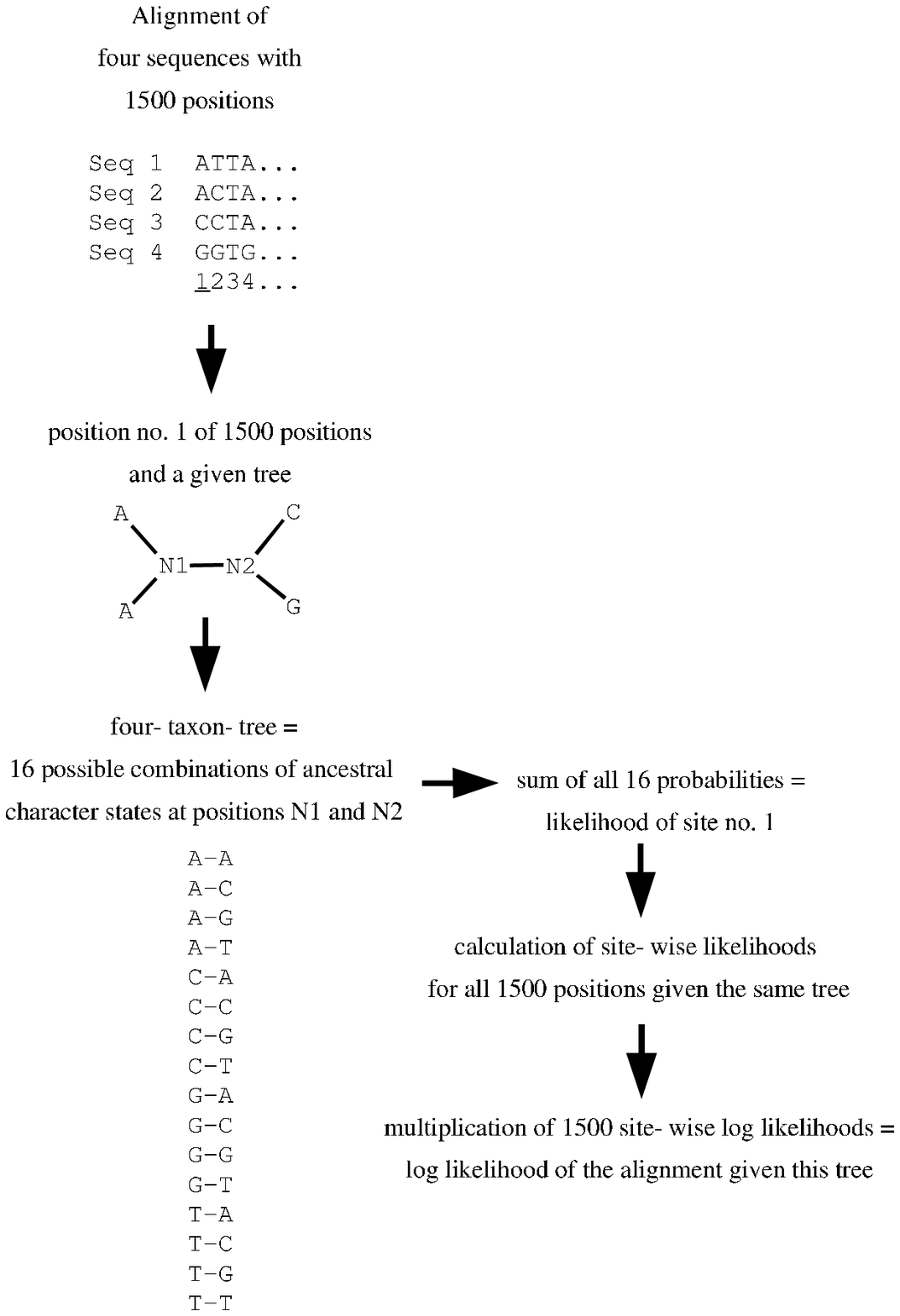}
  \caption{Computation of the likelihood of a tree. To obtain the overall
likelihood value of a tree, for each position of the alignment the
probabilities of all possible combinations of ancestral character states are
computed. The site-wise likelihood comprises the sum of all probabilities.
The site-wise log likelihoods are then multiplied and result in the log
likelihood of a given tree.}
 \end{center}
\end{figure}
It searches for the tree that optimizes the probability of observing the
data. The likelihood of a tree is expressed as negative natural logarithm.
The maximum likelihood method also allows for different evolutionary models,
but differs from distance matrix methods in that it uses discrete characters
and may result in more than one optimal tree (however, rarely more than
two).

The numbers of sequences used to infer phylogenetic trees in biological
research projects almost always prohibited exhaustive searches of the
complete tree space due to limitations of computation time. Thus, maximum
parsimony or maximum likelihood were usually combined with heuristic tree
search algorithms. For a heuristic search a first tree is generated e.g.\ by
adding the sequences step-by-step to the growing tree. This first tree is
then subjected to local and/or global rearrangements by swapping internal
branches or cutting the tree into pieces and rejoining the parts in
different places. This procedure is supposed to overcome potential local
optima and to find the global optimum. The construction of a tree by
neighbor-joining, the preferred method used with distance matrices, starts
with a star-like tree. The pair of sequences with the lowest genetic
divergence is joined (i.e.\ they are said to be neighbors) and the distance
matrix recalculated. These steps are repeated with the next closest related
sequences or clusters of sequences until the tree is completely resolved.

In Bayesian analyses, posterior probabilities for trees and evolutionary
parameters are calculated using the Bayes theorem
\cite{HuelsenbeckEtAl2001}. With the Bayes formula the posterior probability
of a tree given the data is calculated using prior probabilities of the data
and the tree, and the likelihood of a tree. Since it is impossible to
calculate all trees and evolutionary parameters from the space of the joint
posterior probability distribution, samples are drawn using
Metropolis-coupled Markov chain Monte Carlo simulations. This means, at
start of a Bayesian analysis, several chains are initialized to search for
the global optimum in the space of the joint posterior probability
distribution. Once initialized, the chains cross the space for several
hundredthousands to millions of generations by slightly modifying the
parameters (tree topology, branch lengths, evolutionary model parameters).
Trees and evolutionary model parameters are sampled only from the cold
chain; the other so-called heated chains traverse the space more easily and
exchange their status data from time to time with the cold chain. By doing
so, the heated chains help the cold chain to reach the global optimum, which
comprises a set of the best trees and evolutionary parameters. The presumed
global optimum is found when the likelihoods of the trees sampled from the
cold chain reach stationarity.

The phylogenetic trees inferred by the above mentioned methods are usually
bifurcating trees. They may be rooted or unrooted. In rooted trees, the
closest related sistergroup is used to define the direction of evolution in
the sequences. To e.g.\ examine the relationships among chimpanzee, gorilla
and man, the orangutan would be the appropriate outgroup. Unrooted trees are
like looking onto the treetop from above without knowing where the stem is.
In unrooted trees it is not possible to tell, where evolution started and in
which direction the sequences evolve.

\section{Models of Molecular Evolution}

In addition to exponentially growing numbers of possible trees,
phylogenetic analyses are further complicated by the fact that substitution
rates of nucleotides or amino acids may vary. Evolutionary models are an
attempt to approximate the complexity of molecular evolution as close as
possible.

The proportions of the four nucleotides in a DNA sequence may differ from
gene to gene and, thus, need to be considered in phylogenetic analyses (base
frequencies). To account for differing substitution rates for the six types
of point mutations, a substitution rate matrix is used (Fig.\ 3A).
\begin{figure}[htbp]
 \begin{center}
  \includegraphics{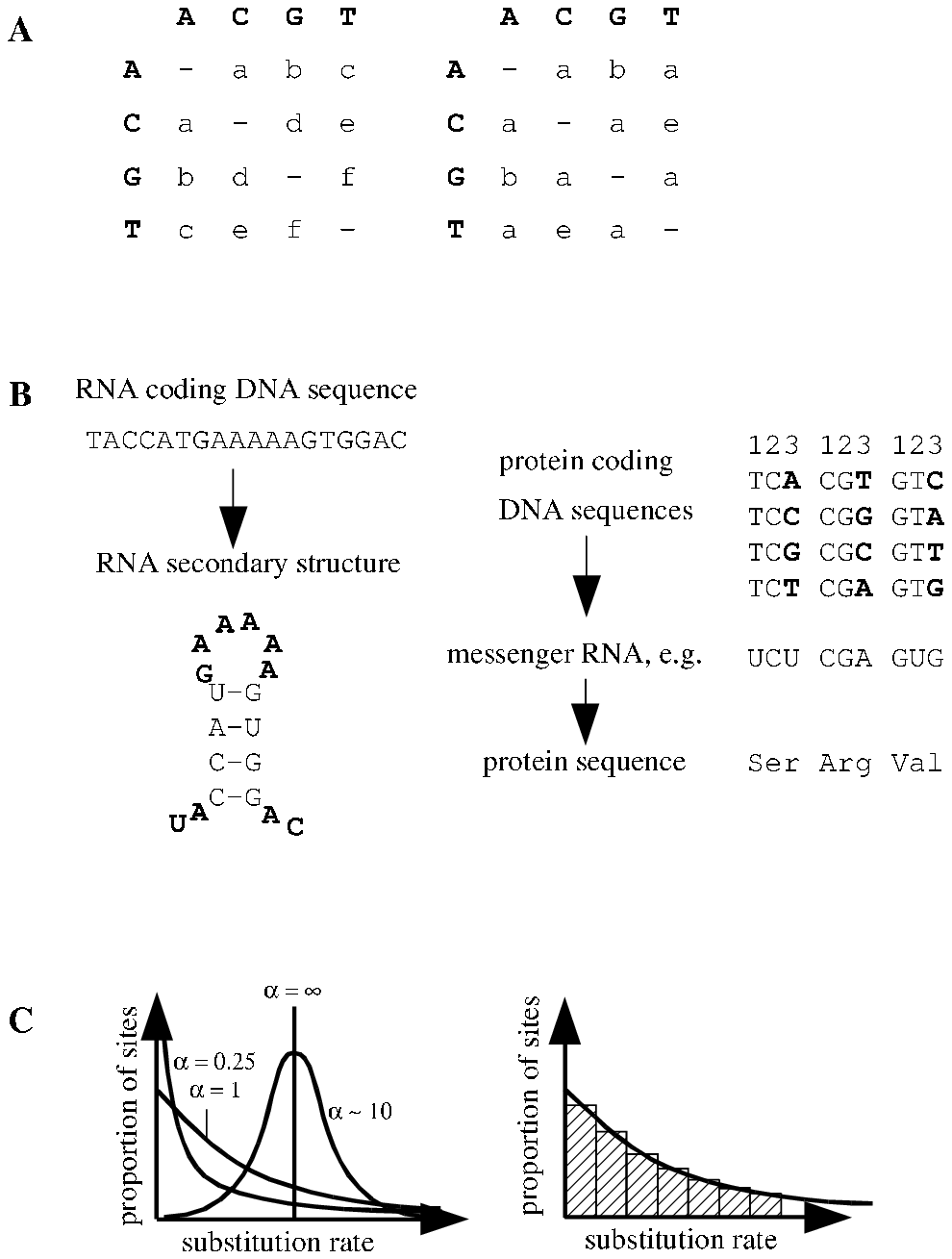}
  \caption{Substitution rate matrices and among-site rate variation. Fig.\ 3A.
Examples for substitution rate matrices. To the left, the most complex type
implemented in phylogeny software programmes, the general time reversible
model (GTR) with six different substitution rates. To the right, a modified
GTR model, the Tamura-Nei model with three different mutation rates. Fig.\
3A. Among-site rate variation in RNA and protein coding DNA. Sites with high
mutation rates are usually found in loop regions of RNA secondary structure,
whereas helices are more conserved (left). In protein coding DNA, the third
position of the codons is usually the most variable. The degenerate code
allows for several codons to represent the same amino acid. In this example,
codons for the amino acids serine, arginine and valine are shown. Between
DNA and protein, a transcription step to messenger RNA is necessary. Bold
face, positions with higher mutation rates. Fig.\ 3C. Modelling the
among-site rate variation using a gamma distribution. Examples for
continuous gamma distribution with different shape parameters to the left
and a discrete gamma distribution with seven rate categories to the right.
The discrete gamma distribution approximates a continuous gamma distribution
with a shape parameter \( \alpha \) of 1.}
 \end{center}
\end{figure}
However, depending in the positions in the alignment, these rates may be
higher or lower. Some positions are highly conserved and do not change at
all. Others evolve at differing rates (Fig.\ 3B). Both parameters, the
proportion of invariable sites and site-specific rate variation, modelled as
a gamma-distribution (Fig.\ 3C), belong to the among-site substitution rate
variation and can be explained by functional constraints on the gene
products.

For most data sets used in biological studies, it is impossible to infer
phylogenetic trees in a reasonable time by optimizing all likelihood
parameters at once during a maximum likelihood analysis, i.e.\ tree topology,
branch lengths of the trees, base frequencies, substitution rate matrix,
proportion of invariable sites and continuously gamma-distributed among-site
rate variation. An often practised approach consisted of determining first
the parameters of the evolutionary model fitting best the data
\cite{PosadaCrandall1998}. To find the appropriate evolutionary model, a
tree is inferred with a fast method (usually distance matrix with
neighbor-joining) and the likelihood values for this tree are calculated for
each available evolutionary model. The model fitting best the data is then
chosen by e.g.\ hierarchical likelihood ratio tests (hLRT) or by the Akaike
information criterion (AIC). Also, a discrete instead of a continuous
gamma-distributed among-site rate variation is used to reduce computation
times (Fig.\ 3C). Thus, during heuristic tree search only tree topology and
branch lengths need to be optimized, whereas the evolutionary model
parameters have been already estimated from the data set using an
approximate tree topology prior to the heuristic tree search.

An additional evolutionary parameter, the covarion/covariotide model takes
lineage-specific evolutionary rates into consideration, i.e.\ complete
sequences may evolve faster than others. The covarion/covariotide model,
however, until today was only implemented in Bayesian phylogenetic analyses
programmes.

Protein coding DNA sequences are \textit{in vivo} first transcribed into
messenger RNA, then translated into a protein consisting of a string of
amino acids (Fig.\ 3B). The function of the protein is determined by folding
up into tertiary and quarternary structures and by amino acids with specific
chemical properties in specific positions. Maximum likelihood analyses of
DNA sequences are quite time intensive. Maximum likelihood analyses with 20
character states for the amino acids are even more time-consuming. Thus, in
protein phylogenies, substitution rate matrices were usually not computed
from the data sets, instead pre-defined sustitution rate matrices
empirically derived from large alignments of other proteins were used
\cite{Felsenstein2003}.

Phylogenetic trees can also be inferred from the DNA sequences of protein
coding genes, which however offers some pitfalls. In protein coding genes,
three nucleotides code for one amino acid, but the genetic code is
degenerate. This means that several three-nucleotide combinations may code
for the same amino acid (e.g.\ six codons are known to code for arginine,
leucine or serine; see Fig.\ 3B). As a consequence, a nucleotide change in
one codon position may be either without effect on the amino acid (= silent
or synonymous substitution), or cause a change of one amino acid to another
(= nonsynonymous substitution). Only nonsynonymous substitutions can result
in a loss or decrease of function, and, thus are subject to functional
constraints. However, the sophisticated evolutionary model parameters
mentioned above were in first place developed to cope with RNA coding genes.
The three-nucleotide codon structure is ignored and synonymous and
nonsynonymous mutations are treated equally. Also, often several
evolutionary pathways are possible to evolve from one codon to another,
which further complicates the evolutionary model parameters. Often the third
positions of codons show nucleotide biases towards higher GC or AT contents.

However, from theoretical and simulation studies, but also empirically, it
became obvious that using wrong assumptions about the underlying
evolutionary processes may result in biased phylogenetic trees.
 
\section{Simulation Studies}

The accuracy of a method comprises consistency, efficiency and robustness. A
method is consistent, if it infers the correct phylogenetic tree with an
infinite amount of data. Efficiency describes the sensitivity of a method
concerning the lengths of sequences. The shorter the sequences can be for a
method to converge to the correct tree topology, the more efficient is the
method. Robustness considers using wrong assumptions about the underlying
evolutionary model. A method is robust, if it infers the correct
phylogenetic tree although a wrong evolutionary model was used. Since
biologists use DNA or protein sequences of finite lengths, in practice only
consistency and robustness of a method are of interest.

In a simulation study by Huelsenbeck \cite{Huelsenbeck1995}, e.g.\ four-taxon
data sets of differing sequence lengths were generated \textit{in silico}
from a random starting sequence according to pre-specified evolutionary
models and phylogenetic trees (see parameter space in Fig.\ 4A).
\begin{figure}[htbp]
 \begin{center}
  \includegraphics{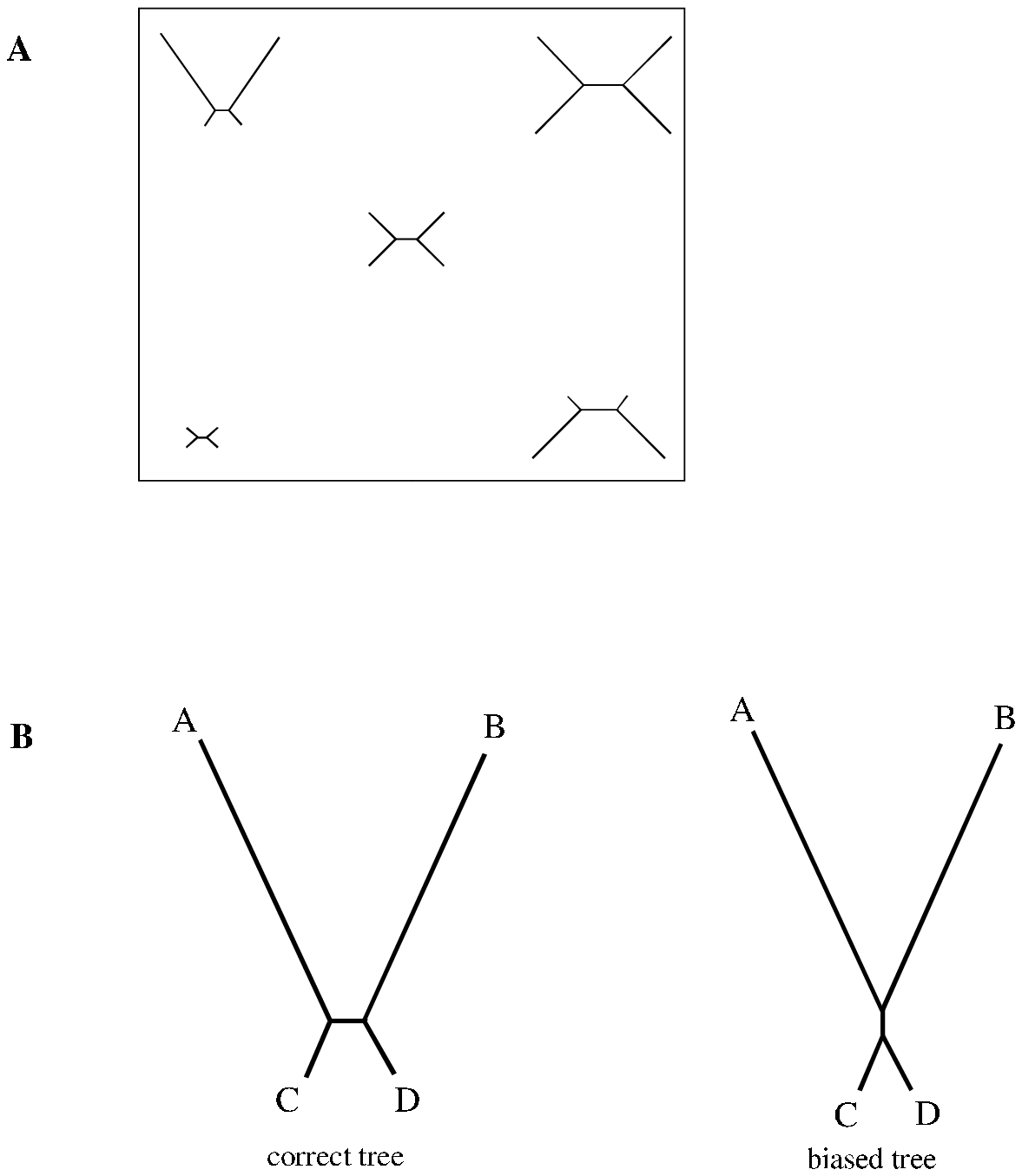}
  \caption{The long branch attraction artefact (LBA). Fig.\ 4A. The parameter
space with different tree topologies usually used in simulation studies with
four-taxon trees. Fig.\ 4B. An example for a LBA of a four-taxon tree. The
tree to the left corresponds to the tree in the top left corner of the
parameter space in Fig.\ 4A. To tree to the right shows the typical LBA bias.
The high evolutionary rates displayed by the long branches of the taxa A and
B cause reversals in the nucleotides, e.g.\ a C mutates to a G, a T and back
to a C. In combination with a high background noise, which blurs
phylogenetic signals, these reversals are presumably interpreted erroneously
as positives for genetic relatedness. The region in the parameter space
resulting in biased trees is also sometimes called the ``Felsenstein" zone
of a method. This region is predominantly located in the top left, sometimes
extended to the top right of the parameter space shown in Fig.\ 4A. The
larger this ``Felsenstein" zone is, the less robust the phylogenetic
method.}
  \end{center}
\end{figure} 
Different phylogenetic analyses methods were then used to infer trees from
the data sets and the conditions determined that caused the methods to infer
wrong tree topologies. The so-called long branch attraction artefact (LBA)
is the most well-known phenomenon causing biased tree topologies. Usually,
LBAs were found in phylogenetic trees with extremely long terminal (i.e.\
branches with high evolutionary rates) but short internal branches (Fig.\
4B). In most test situations, maximum likelihood outperformed other methods,
but it also failed in finding the correct tree, if the assumed evolutionary
models were too different from the evolutionary processes under which the
simulated data sets had evolved.

\section{Phylogenetic Analyses and Real Life Data}

Since divergent branch lengths were almost always found in phylogenetic
analyses of \textit{in vivo} evolved sequences, the effects of potential
LBAs were a frequent matter of concern \cite{AndersonSwofford2004}.
Especially in large scale phylogenies comprising sequences of very different
organisms, long-branch taxa were often gathered ladder-like close to the
root of the trees, which may indicate a potential bias caused by LBAs. The
farther back in time the examined relationships of organisms reach, the
worse the resolution at the internal branches of a tree. It was found that
an addition of sequences to the data set and a complex evolutionary model
with a gamma-distributed among-site rate variation were the best options to
reduce artefacts in a phylogenetic tree \cite{Graybeal1998},
\cite{BrunoHalpern1999}. Especially, adding more sequences of the
problematic type could break up long branches, increase the resolution in
this part of a tree and thereby neutralise the LBA.

An example of how taxon sampling and choice of evolutionary model may affect
the results of a molecular phylogeny can be found in the cryptophytes, a
group consisting of microscopic flagellated unicells. Most of the genera in
this group are algae, i.e.\ they contain a pigmented plastid which is used to
turn the energy of light into chemical energy by photosynthesis. Two genera
are, however, colourless. \textit{Goniomonas} is phagotrophic; it feeds from
ingesting bacteria. The other genus, formerly classified as
\textit{Chilomonas} feeds from organic molecules, but still harbours a
leukoplast, i.e.\ a colourless plastid. In a phylogenetic analysis with a low
number of nuclear 18S ribosomal DNA sequences, \textit{Goniomonas} and
``\textit{Chilomonas}" clustered together indicating a relationship of both
genera \cite{CavalierSmithEtAl1996}. In a later analysis, sequences of the
photosynthetic genus \textit{Cryptomonas} were added \cite{MarinEtAl1998}.
It turned out that \textit{Goniomonas} was the most basally diverging taxon,
whereas ``\textit{Chilomonas}" was a colourless \textit{Cryptomonas}. The
clade with the genera \textit{Cryptomonas} and ``\textit{Chilomonas}" seemed
to be the most basal group of the plastid-bearing cryptophytes. Thus, the
sisterhood of \textit{Goniomonas} and ``\textit{Chilomonas}" were caused by
a LBA due to inappropriate taxon sampling. The analysis in
\cite{MarinEtAl1998}, however, was done using maximum likelihood under a
simple evolutionary model, i.e.\ without considering an among-site rate
variation. In a study using a complex evolutionary model with among-site
rate variation, the basal position of the
\textit{Cryptomonas}/``\textit{Chilomonas}" clade was also shown to be an
artefact caused by long branch attraction \cite{HoefEmdenEtAl2002}.

Thus, long branch attraction artefacts are a real problem in phylogenies
inferred from \textit{in vivo} evolved sequences. The best options to cope
with LBAs, i.e.\ adding more taxa, and using complex evolutionary models and
robust methods, however, collide with another problem biologists were and
are still confronted with computation times. The larger the amount of
sequences, the more reliable the phylogenetic analyses methods do work, but
exponentially more time is also needed to obtain results.

Bayesian analysis was introduced as a potential faster alternative to
maximum likelihood analysis \cite{HuelsenbeckEtAl2001}. However, for large
data sets Markov chains often need to be run for more generations to reach a
plateau of likelihood values, which also increases comutation times. In
addition, the posterior probabilities given for the different branches of
the consensus tree, in which the sampled trees are summarised, are more
optimistic than support values obtained from nonparametric bootstrapping
using the maximum likelihood criterion (i.e.\ a subsampling method with at
least 100, often more than 100 subsample data sets, to test the stability of
the branches of a tree). Bayesian analysis may be speeded up by running the
different Markov chains on separate CPUs of a computing server or a cluster.

In heuristic tree searches using the maximum likelihood criterion, some
parallelised versions of programmes have been introduced e.g.\
\cite{StewartEtAl2001}. The tasks of tree generation and tree evaluation
were distributed among a master (tree generation and comparison) and worker
programmes (calculation of branch lengths and likelihoods).

Another attempt to decrease computation times was quartet-puzzling
\cite{StrimmervonHaeseler1996}. In quartet-puzzling, trees are computed from
quartets of n sequences of a larger data set using the maximum likelihood
criterion and weighted accordingly. The best of the three possible 4-trees
for each quartet are used to first assemble a large number of n-trees
(quartet-puzzling) and finally to obtain a consensus n-tree. This method is
much faster than a heuristic trees search, but more vulnerable to LBA. Among
hundreds to thousands computed four-taxon trees, only a low number of
biased 4-trees suffices to pass on a topological error to the final n-tree.
In simulation studies, global character maximum likelihood almost always
outperformed quartet-puzzling or related methods \cite{RanwezGascuel2001}.

Other studies tried to overcome LBA and exponentially growing computing
times with longer sequences, e.g.\ by using complete genomes to infer
phylogenetic trees. Phylogenetic analyses of longer sequences increase the
computing times only linearly. Since sequencing of complete genomes need
much more time and resources than that of single genes or smaller sets of
genes, the taxon sampling in these studies generally was lower. It has been
shown, however, that long sequences cannot compensate for an extended taxon
sampling. The low number of taxa included in a genome-scale analysis
resulted in high bootstrap support even for biased tree topologies
\cite{SoltisEtAl2004}. Also genome-scale alignments cannot be refined
by eye anymore. They depend in automatic alignment algorithms, which may
perform badly by producing more or less biologically meaningless alignments
\cite{PollardEtAl2004}. A better option than using complete genomes
presumably is to sequence a set a of genes, to refine the alignment of each
gene by eye, and to concatenate the genes \cite{BaptesteEtAl2002}.

Additional problems occur, if the evolution of a gene and/or a group of
organisms cannot be described by bifurcating trees. In sexually reproducing
populations, the examined gene may be present in differing alleles. Each
individual of a population inherits two alleles, one from its mother, the
other from its father. In addition, parts of the alleles can be exchanged by
genetic recombination. Genetic material may also be transferred between
unrelated organisms, e.g.\ by infection with viruses, by endosymbiosis or in
bacteria by exchange of plasmids. Whereas the inheritance of genes from
parents to child is called vertical gene transfer, the exchange of genetic
material between unrelated organisms is called lateral gene transfer. The
results of sexual reproduction or lateral gene transfers are genetic
chimaeras and reticulate evolutionary trees.

\section{Conclusions}

Until yet, there seems to be no easy way out of the treadmill of extremely
increasing computing times for phylogeneticists. New algorithms to reduce
time consumption in phylogenetic analysis have been proposed until recently,
e.g.\ \cite{GuindonGascuel2003}. However, only if the algorithms are offered
in software programmes suitable for the tasks of phylogenetic analysis, if
they are presented in an understandable way to biologists and if they
prove to be robust, they will accepted and used.

\bigskip

\bibliographystyle{plain}

\end{document}